\documentclass{agujournal}
\draftfalse

\journalname{Geophysical Research Letters}

\begin{document}

%
%


\title{On the Close Correspondence between Storm-time ULF Wave Power 
and the POES VLF Chorus Wave Amplitude Proxy}

%
%




 \authors{Stavros Dimitrakoudis\affil{1}, 
Ian R. Mann\affil{1}}

\affiliation{1}{Department of Physics, University of Alberta, Edmonton, Alberta, Canada}




\correspondingauthor{S. Dimitrakoudis}{dimitrak@ualberta.ca}




\begin{keypoints}
\item There is a common L-shell and time profile of storm-time ground-based 
ULF wave power and the POES proxy for VLF wave amplitude
\item Global ground-based ULF wave power coherence implies a small number of 
meridians can be 
used to estimate radial diffusion coefficients 
\item ULF wave power and the POES VLF wave proxy discriminate identically between efficient 
and inefficient radiation belt acceleration events 
\end{keypoints}

%
%


\begin{abstract}
Ground-based Pc5 ULF wave power in multiple ground-based meridians is compared to the VLF 
wave amplitude proxy, derived from POES precipitation, for the 33 storms studied by \cite{Li2015}.
The results reveal common L-shell and time profiles for the ULF waves and VLF proxy 
for every single storm, especially at $L\leq 6$, and identical discrimination between efficient and
inefficient radiation belt electron acceleration. The observations imply either ULF waves play 
a role in driving precipitation which is falsely interpreted as VLF wave power in the proxy, 
ULF waves drive VLF waves (the reverse being energetically unfeasible), or both have a common 
driver with nearly identical L-shell and time-dependence. Global ground-based ULF wave 
power coherence implies a small number of meridians can be used to estimate 
storm-time radial diffusion 
coefficients. However, the strong correspondence between ULF wave power and VLF wave 
proxy complicates causative assessments of electron acceleration.
\end{abstract}

%
%

%


%
%
%
%

\section{Introduction}

The Earth's outer radiation belt is populated mostly by electrons, which
are at times accelerated to relativistic energies ($>1$MeV), and can thus pose
a danger to satellites and astronauts \citep{Baker1998, Webb2004, Daglis2004,
Choi2011, Sarno-Smith2016}. 
It is therefore important to determine the complete set of processes
by which this acceleration takes place, starting with solar wind
drivers and ending with changes to electron dynamics near Earth, for
the purpose of forecasting and risk mitigation. Two acceleration mechanisms 
have emerged as favoured candidates for the dominant role in the process: 
energy diffusion by chorus waves  \citep{
Horne1998, Summers2002, Horne2005, Tao2009, Reeves2013, Thorne2013, Li2014a,
Tu2014, Li2016}, or inward radial diffusion by ultralow frequency (ULF) waves 
\citep{Falthammar1965, 
SchulzLanzerotti1974, Elkington1999, Hudson1999, Mathie2000, Perry2005, 
Fei2006, Ukhorskiy2009, Huang2010, Turner2012, Mann2016, Cunningham2016}.

\cite{Li2015} (henceforth L15) recently applied an indirect method of 
calculating chorus wave intensity using a VLF wave proxy defined using 
POES electron precipitation measurements  \citep{Li2013POES, Ni2014POES} 
to examine the potential role of VLF waves in radiation belt dynamics. 
L15 compared this VLF proxy to the response of the outer electron radiation 
belt for 33 storms, which they separated into 16 efficient 
acceleration (EA) and 17 inefficient acceleration (IA) events based on 
the relativistic electron response, and found
a correlation between the VLF chorus wave proxy and the efficiency of 
electron acceleration. Since the L15 analysis was only performed using the VLF 
chorus wave proxy, the question of how well ULF waves correlate with those 
same events was left open. It was indirectly revisited by 
\cite{Li2016}, where diffusion coefficients from chorus waves, ULF radial diffusion, 
and plasmaspheric hiss were calculated for the 17 March 2015 geomagnetic storm; 
but the radial diffusion coefficients were derived from analytical expressions  
\citep{Brautigam2000, Ozeke2014}, which are statistically derived, 
and may not always perfectly correspond to those in each particular event 
\citep{Murphy2016}.  
Here we revisit the events considered by L15 using a new method of visualizing ULF wave 
power across multiple L-shells, and in multiple MLT meridians, as a function of time by 
using data from multiple ground magnetometer chains deployed across the Earth's northern 
hemisphere. The large amount of ULF wave data available from these arrays allows us to 
perform a ULF wave superposition analysis much like L15 did for chorus waves, and to hence 
directly compare both the distributions of the ULF wave power and the VLF chorus wave proxy, 
and their potential role in radiation belt electron acceleration.

We present our methodology in Section~\ref{methodology}, our results in 
Section~\ref{results}, and conclude with a discussion in 
Section~\ref{discussion}.

\section{Methodology}
\label{methodology}

We use the same 16 efficient acceleration (EA) and 17 inefficient acceleration (IA) 
electron radiation belt events identified by L15 during the period from October 2012 
to March 2015. We calculated the integrated Pc5 ULF power every hour for seven days 
spanning the minimum in electron flux for each one of those storm events. We then 
superposed the ULF wave data for the EA and IA events separately, in order to produce 
figures that can be directly comparable to both the VLF chorus wave proxy for each event, 
and the superposed epoch analysis of the VLF proxy for the EA and IA events, presented in L15. 
The same VLF chorus wave amplitude proxy derived by L15 is used in this study, where 
the ratio of POES-observed 30-100 keV electron loss-cone to trapped fluxes averaged 
over all MLT regions is used 
to generate a global proxy for VLF wave chorus amplitudes.  
For direct comparison, in the ULF wave power figures presented below we also show 
the relevant VLF chorus wave proxy data from L15 in the last panel. 

For all events the zero epoch was defined in the same way as L15, being the time 
when the maximum electron phase space density (PSDmax), taken over the range of $2.5 - 6 \mathrm{R_E}$
and at first adiabatic invariant ($\mu = 3433 \mathrm{MeV/G}$) and second adiabatic 
invariant ($K = 0.10 G^{1/2} \mathrm{R_E}$) as measured by the Van Allen Probes Relativistic 
Electron Proton Telescope (REPT) instrument \citep{Baker2013REPT, Spence2013RBSP} reaches a minimum. 
Note also that L15 defined an EA (IA) event  
if within two days prior to the zero epoch time PSDmax decreased at least by a factor of 5,
while within two days after that it was larger (smaller) than $10^{-8} (10^{-9}) c^3\mathrm{ MeV}^{-3} \mathrm{cm}^{-3}$. This is also the categorization used here. 

\begin{figure*}
\noindent\includegraphics[width=35pc]{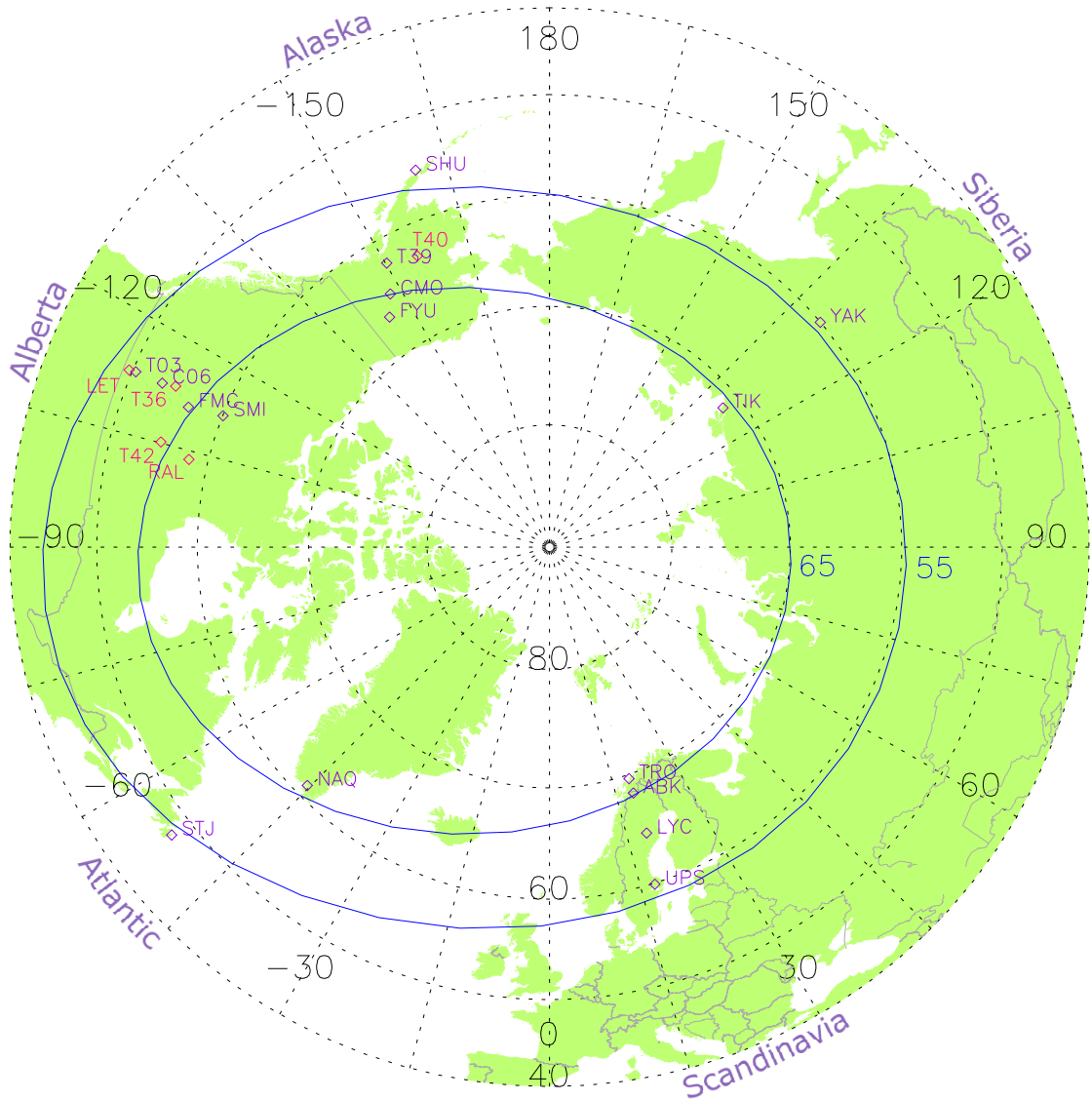}
\caption{Map of the selected ground magnetometer stations used in this study (cf. Table~\ref{table1}). 
The stations belonging to the standard set are in purple, while additional temporary replacement 
stations are shown in magenta. 
The blue curves are magnetic latitude contours, at $55^\circ$N and $65^\circ$N in 2013.}
\label{map}
\end{figure*}

To build a comprehensive picture of the global nature of ULF wave power as a function of L, 
MLT and time for these 33 events, we used measurements from 21 selected ground magnetometer 
stations available from the SuperMAG archive, which offers easy access to standardized data 
from over 300 ground magnetometers from multiple arrays from around the world 
\citep{Gjerloev2009, Gjerloev2012}. 
More specifically, we focused on data from five latitudinal chains of northern stations 
from different longitude sectors, labelled as Atlantic, Alberta (Canada), Alaska, Siberia, 
and Scandinavia, at latitudes that correspond to invariant dipole L-shells from $L \sim 2.5$ to 
$L \sim 6.5$ (cf. Figure~\ref{map}). For most of the selected stations, data was available for 
all of the 33 events. However, a small number of the selected stations in key L-shell 
and longitudinal sectors had some data gaps and in such cases alternative nearby stations 
were used where possible to try to fill in the coverage.
Data from the selected stations are all shown in  Figure~\ref{map} 
and their names and coordinates are displayed in Table~\ref{table1}. 
The SuperMAG data used here was standardized to a 1-minute cadence, so by performing 
Fast Fourier Transform (FFT) as described in \cite{Rae2012}
we were able to obtain power spectra in the Pc5 frequency range, from 1.68 to 7mHz, spaced with 
a uniform linear step of 0.28 mHz. These Pc5 power spectra were derived for each hour, from all 
of the available selected stations, and their summed power from 1.68 to 7mHz were then plotted 
in bins of width of one dipole L-shell.

\begin{table}
\caption{Ground magnetometer stations used in this work. }
\centering
\begin{tabular}{ l l c c c c c }
 \hline 
 Code & Name & Geo. Lat. / Lon. & 
 CGM Lat. / Lon. & 
 L-shell & primary  \\
  & & $(^{\circ})$N / $(^{\circ})$E & $(^{\circ})$N / $(^{\circ})$E & & operator \\ 
 \hline
Alaska \cr 
 \hline   
FYU	& Fort Yukon & 66.56 /  214.79 & 67.65 /  267.17 & 6.85 & GIMA \cr
CMO	& College & 64.87 / 212.14 & 65.45 / 266.18 & 5.77 & GIMA \cr
T39	& Trapper Creek & 62.30 / 209.80 & 62.37 / 265.76 & 4.65 & GIMA \cr
SHU	& Shumagin & 55.38 / 199.54 & 53.29 / 260.26 & 2.83 & USGS \cr
 \hline 
Alberta \cr 
 \hline  					
SMI	& Fort Smith & 60.03	/ 248.07 & 67.47 / 307.71 & 6.74 & CARISMA \cr
FMC	& Fort McMurray & 56.66	/ 248.79 & 64.28 / 309.98 & 5.3 & CARISMA \cr
C06	& Ministik Lake & 53.35	/ 247.03 & 60.64 / 308.76 & 4.19 & CARISMA \cr
T03	& Vulcan & 50.37	/ 247.02 & 57.60 / 309.60 & 3.53 & CARISMA \cr
 \hline 
Atlantic \cr 
 \hline 					
NAQ	& Narssarssuaq & 61.16	/ 314.56 & 65.75 / 43.19 & 5.8 & DTUSpace \cr
STJ	& St Johns & 47.60	/ 307.32 & 52.60 / 31.64 & 2.72 & CANMOS \cr
 \hline 
Scandinavia \cr 
 \hline 					
TRO	& Tromso & 69.66	/ 18.94 & 67.07 / 102.77 & 6.56 & IMAGE \cr
ABK	& Abisko & 68.35	/ 18.82 & 65.74 / 101.70 & 5.9 & IMAGE \cr
LYC	& Lycksele & 64.61	/ 18.75 & 61.87 / 99.33 & 4.49 & IMAGE \cr
UPS	& Uppsala & 59.90	/ 17.35 & 56.88 / 95.95 & 3.35 & IMAGE \cr
 \hline 
Siberia \cr 
 \hline 					
TIK	& Tixie & 71.59	/ 128.78 & 66.70 / 198.71 & 6.4 & AARI \cr
YAK	& Yakutsk & 60.02	/ 129.72 & 54.88 / 202.60 & 3.05 & SHICRA SB RAS  \cr
 	&   &   &   &   & \& GFZ  \cr
 \hline
Additional 
stations\cr 
 \hline   					
T40	& McGrath & 63.00	/ 204.40 & 62.16 / 260.84 & 4.59 & THEMIS \cr
RAL	& Rabbit Lake & 58.22	/ 256.32 & 67.00 / 319.92 & 6.46 & CARISMA\cr
T42	& La Ronge & 55.15	/ 254.84 & 63.76 / 318.65 & 5.1 & AUTUMN \cr
T36	& Athabasca & 54.71	/ 246.69 & 61.95 / 307.91 & 4.54 & AUTUMN \cr
LET	& Lethbridge & 49.64	/ 247.13 & 56.88 / 309.93 & 3.39 & AUTUMN \cr
 \hline
\label{table1} 
\end{tabular}
\end{table}

\section{Results}
\label{results}

\begin{figure*}
\noindent\includegraphics[width=40pc]{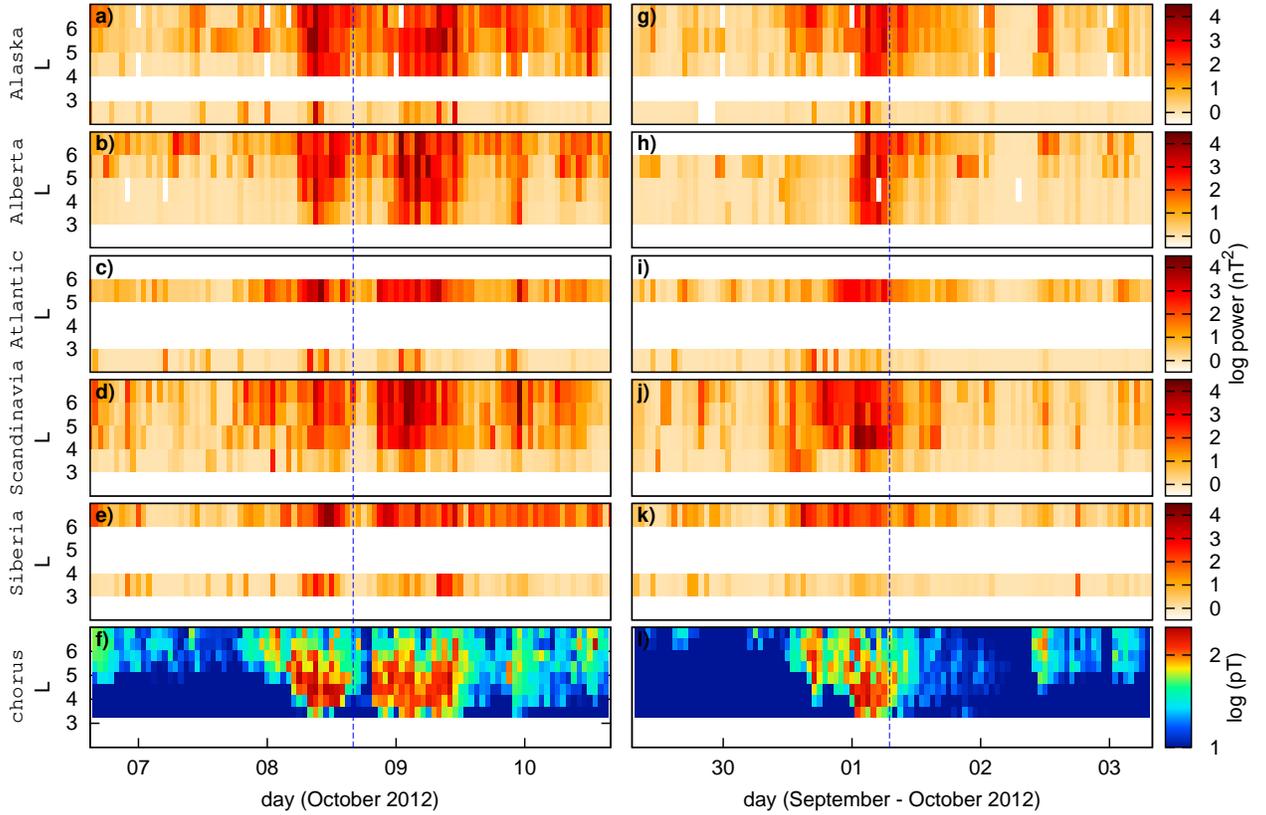}
\caption{(panels a-e) ULF wave power for a representative  efficient acceleration (EA) event 
during the 8-9 October 2012 storm, each panel showing data as a function of L-shells in five 
different longitudinal meridians. (panel f) VLF chorus wave amplitude proxy, averaged over 
all MLT sectors, from L15. The blue line indicates the zero epoch time (see text for details).
Panels (g-l) as in panels (a-f) but for a representative inefficient acceleration (IA) 
event on 1 October 2012.}
\label{ea1}
\end{figure*}

Panels a-f in Figure~\ref{ea1} show ULF power for a 
representative EA event on 8-9 October 2012, which can be 
compared directly to Figure 1 in L15. We see an extremely strong similarity of the ULF wave power 
distributions with the L15 POES VLF chorus wave proxy, replotted here in panel f. 
Both L-shell and time profiles of the ULF wave power with that of the POES VLF chorus 
are seen to enhance in the day before the minimum flux at epoch time,
with each gradually penetrating to lower L-shells during the time that radiation belt flux
is decreasing up to epoch time zero 
(cf. Figure 1 of L15). The subsequent times after epoch time zero, during which the 
efficient radiation belt acceleration takes place, are accompanied by ongoing enhancements 
in both ULF wave power and in the POES proxy for VLF chorus wave amplitudes which last for 
days.  
What is also immediately apparent is the clear global temporal coherence of the enhancements 
in ULF wave power spanning all longitudinal meridians, and the extremely close correspondence 
of these globally coherent ULF wave power enhancements (panels a-e) and those of 
the global VLF chorus wave amplitude proxy from L15 (panel f).

For comparison, panels g-l show the ULF wave power distributions and the VLF 
chorus wave amplitude proxy for a characteristic IA event on 1 October 2012 and which can 
be compared directly to Figure 2 in L15. Once again, there is an obvious and striking 
similarity between the L-shell and time profiles of the ULF wave power in every longitudinal 
meridian and the VLF chorus wave amplitude proxy. The characteristic ULF and VLF proxy 
profiles before epoch time zero for this IA event also look very similar to those for the 
EA event shown in panels a-f. However, and in contrast to the ULF and VLF characteristics 
for the EA event in panels a-f, for the IA event in Figure 3 the post epoch time zero ULF 
and VLF wave power both essentially totally disappear. The only significant difference 
between the L-shell and time profiles of the ULF wave power and the chorus proxy in all of the 
panels is the relative lack of activity in the chorus wave proxy at $L > 6$, 
compared to ULF power which continues to be enhanced at these higher L-shells at the same time. 
Most likely this can be explained by the relative lack of the equatorial radiation belt electron 
source population required for electrons to be precipitated and to generate a signal in the 
POES VLF chorus wave proxy, or alternatively a population of higher L-shell ULF waves 
which are not associated with POES precipitation. We have completed the same analysis for every one of the remaining 
31 storms in the L15 set, and these plots are provided in the Supplementary Material. 
The extremely close correspondence between the L-shell and time envelopes of the ULF wave power and the 
POES VLF chorus wave amplitude proxy is maintained in essentially every single event,
while a global temporal coherence of the ULF wave power in multiple longitudes is
also seen in most events.

\begin{figure*}
\noindent\includegraphics[width=40pc]{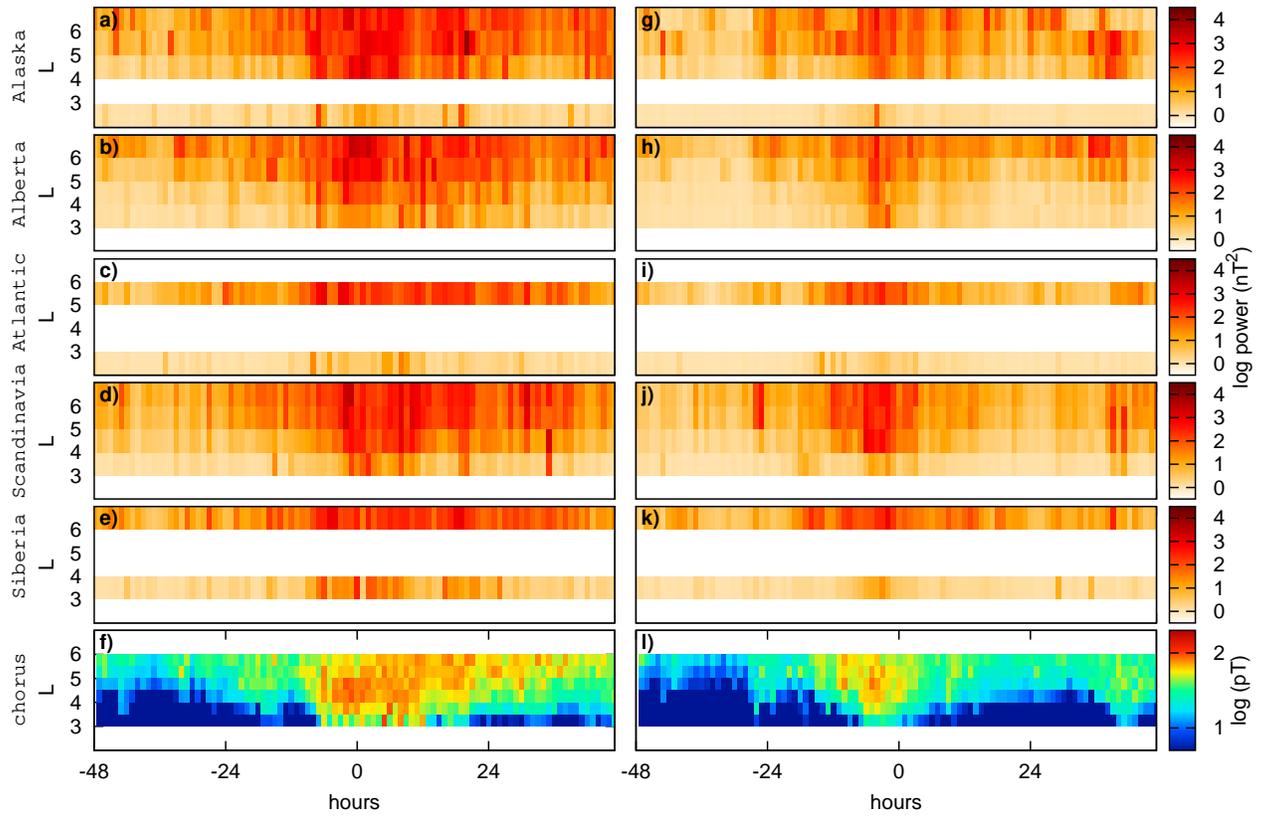}
\caption{Superposed epoch analysis of ULF wave power as a function L-shell and time in 
five different longitudinal meridians (top five rows of panels), and the VLF chorus wave 
proxy from L15 (bottom row of panels), for efficient acceleration (EA, panels a-f) and 
inefficient acceleration (IA, panels g-l) events.}
\label{eas}
\end{figure*}

\begin{figure*}
\noindent\includegraphics[width=40pc]{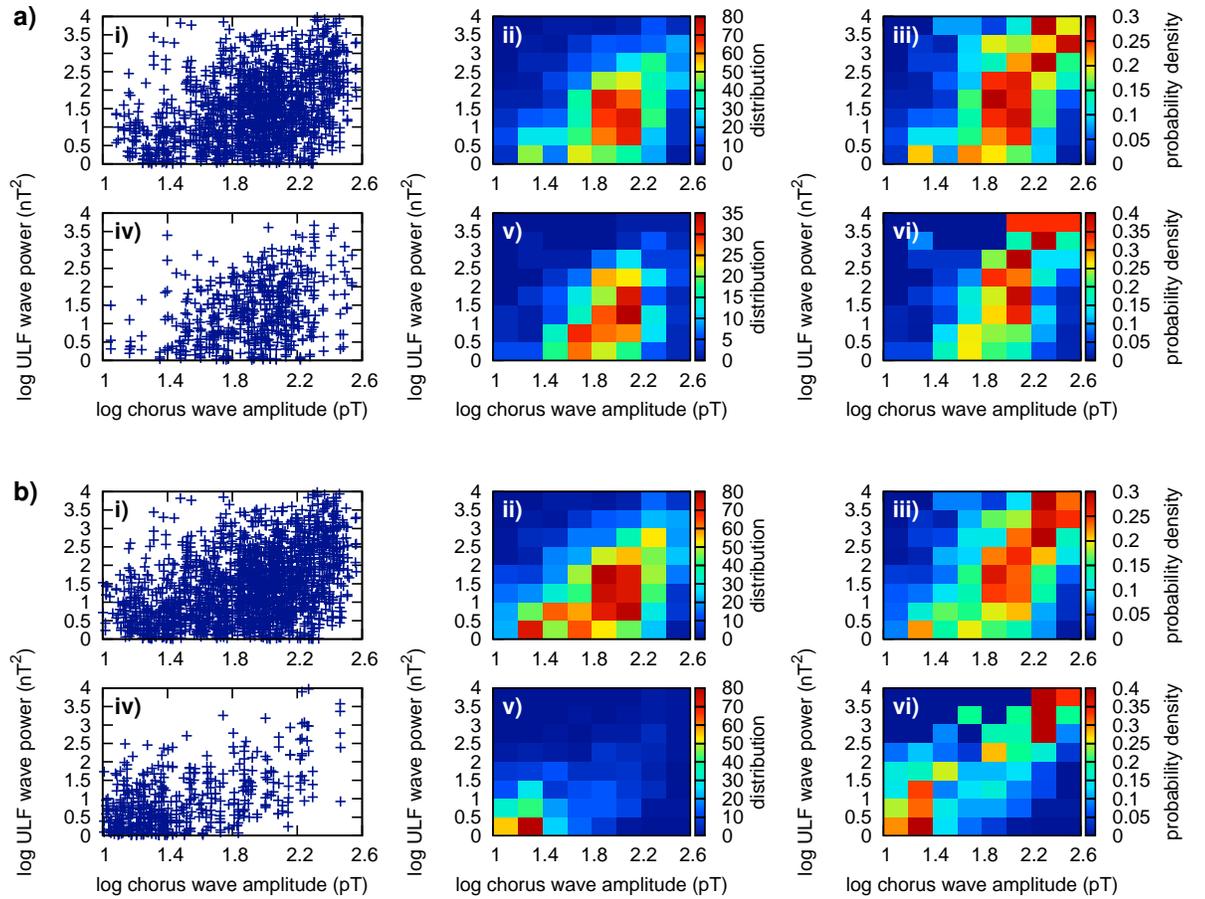}
\caption{a) L = 5 to 6: i) Scatterplot of integrated Pc5 ULF power measurements compared to the
VLF chorus wave amplitude proxy for all EA events;
 ii) histogram of the scatterplot at (i);
iii) probability distribution of VLF chorus wave amplitude proxy as a function of ULF power
measurements for all EA events. iv-vi) as in (i-iii) but for all IA events. b) as in (a) 
but for L = 4 to 5.} 
\label{dist}
\end{figure*}

Following L15, we also completed a superposed epoch analysis for all of the events in the 
EA and IA categories, centered around their respective zero epochs. Figure~\ref{eas}, 
panels (a-e), shows the results from this superposition of ULF wave power in each meridian 
for the EA events, and this should be compared to the superposed epoch analysis of the VLF 
wave proxy shown in Figure~\ref{eas} (f) (reproduced from Figure 3h of L15). Figure~\ref{eas}, 
panels (g-k) shows the results of the same analysis for the IA events, 
and where the superposition of the POES VLF chorus amplitude proxy from Figure~\ref{eas} (l) 
is the same as that shown in Figure 4h of L15. The characteristics of the superposed ULF wave 
power from every longitudinal meridian is almost identical to that of the VLF chorus wave proxy 
in shape in L-shell and time, both for the EA and IA events. Especially for the EA events 
(Figure~\ref{eas}, panels (a-e)) there appears to be little difference in the ULF wave power 
profiles between the meridians in different longitudes, and every meridian has the same 
characteristics as the VLF wave proxy. The same is true for the IA events (Figure~\ref{eas}, 
panels (g-k)), with perhaps a slightly larger ULF power being seen in the Scandinavian meridian 
as compared to the other meridians. Overall, just as in the case of the individual EA and IA 
events (Figure~\ref{ea1}), the superposed epoch ULF wave profiles are almost 
identical to those of the VLF chorus wave amplitude proxy. 
Figure~\ref{dist} shows probability distribution plots for all
EA and IA events, demonstrating that 
high values of integrated ULF power predominantly coincide with high amplitudes of the
VLF chorus wave proxy.

As shown in the
supplementary material, for the bands L = 4 to 5 and L = 5 to 6, Pearson correlation coefficients
between ULF and VLF wave proxy values peak at zero temporal offset, with r = 0.65 to 0.84 and
the p-values for the statistical significance of the null hypothesis 
ranging from $p=3\cdot 10^{-8}$ to $p< 10^{-16}$, using
a 2-tailed test of significance.
Of course the sample size is large, but this indicates that the connection between the 
VLF proxy and ULF wave power was very unlikely to have occurred by chance.
Figure S36 in the Supplementary material illustrates the meridian dependence, showing the ULF wave global 
coherence between Albertan and Scandinavian meridian ULF wave power. 
The fractional variance is consistent in log-log space, and even though
some inherent variability exists, likely stemming from MLT dependence, it is
dwarfed by the coherent variability of 
ULF wave power through the course of the storms.
Therefore, overall, Figure S36 shows an obvious coherence between ULF waves observed in both meridians.

\section{Discussion}
\label{discussion}

A correlation between a VLF wave proxy, derived from POES electron precipitation, and radiation 
belt electron acceleration events has previously been used by L15 to argue VLF waves play an 
essential role. Here we show that both the VLF wave proxy and the ULF wave power have 
common profiles in L-shell and time, not only statistically in relation to their 
characteristics during radiation belt electron efficient acceleration (EA) and inefficient 
acceleration (IA) events, but also in each of the 33 individual storms examined by L15. 
Moreover, the ULF wave power when analysed across multiple meridians tends to show a rather 
remarkable global coherence as a function of MLT, which suggests that data-driven radial 
diffusion coefficients might be able to be derived not only historically but also in 
near-real time using data from stations from a small number of meridians.    

Our analysis
shows that the question of whether ULF or chorus waves are 
responsible for electron acceleration is far from settled, and certainly challenges the 
interpretation of L15 that the VLF waves should be identified as definitively playing a 
key role in electron acceleration. Of course VLF chorus waves might be important, but 
given the results presented here they cannot necessarily be identified as the dominant 
acceleration process based on the characteristics of the VLF proxy during EA and IA events alone. 

The strong and clear correlation between ULF wave power and the L15 chorus proxy invites 
the following possible explanations:

\begin{enumerate}
\item \label{Chorus_first} chorus waves emerge first and cause a growth of ULF waves;
\item \label{ULF_first} ULF waves emerge first and cause a growth of chorus waves;
\item \label{common_cause} both ULF and chorus waves are driven with identical L-shell 
and time profiles by some common solar wind and/or magnetospheric driver which drives 
both in the same place at the same time;
\item \label{ULF_precipitation} ULF waves can have a direct impact on radiation belt 
electron precipitation, which pollutes the POES precipitation proxy which is used to 
infer the presence of VLF chorus waves.
\end{enumerate}

Case~\ref{Chorus_first} is implausible, because chorus waves contain much less energy than ULF waves. 
Case~\ref{ULF_first}, on the other hand, has some observational and theoretical basis
\citep[and references therein]{Li2011}. For example, compressional Pc4-5 pulsations have 
been found to produce conditions that encourage the production or modulation of chorus waves.  
If the reservoir of energy in ULF waves can act 
to locally produce VLF waves, or encourage their more efficient growth, then this would be 
a natural explanation for the observations. However, we are not aware of any processes 
through which ULF waves can directly produce chorus emissions.

Case~\ref{common_cause} is less restraining than the previous two cases, 
since the only correlation needed between the two wave types is temporal, so long as 
the driving processes act to produce both wave modes in the same place. Chorus waves are 
assumed to be excited near the Earth's geomagnetic equator by cyclotron resonant interactions 
with electrons that are anisotropically injected into the plasma sheet during geomagnetically 
active conditions, either preceding a substorm onset in the dawnside \citep{Lyons2005, Hwang2007}, 
or following a substorm onset near midnight \citep{Smith1999}. ULF waves in the 
Pc5 frequency range have also been observed to be more prominent in the dawnside
\citep{Anderson1990, Ruohoniemi1991, Glassmeier2000}, generated either by
shear flow instabilities along the magnetopause 
\citep{Cahill1992, Mann1999}, or more generally in the dayside magnetosphere by variations 
in solar wind dynamic pressure  \citep{Lysak1992}. These studies suggest that the conditions 
which lead to the excitation of ULF and VLF waves are not necessarily the same, and in our view 
it is challenging to explain the observed close correspondence of the ULF waves and the VLF wave 
proxy in terms of a common driver.  

Case~\ref{ULF_precipitation}
would require a mechanism with which ULF waves directly cause electron precipitation, 
thereby being (at least partially) responsible for polluting the chorus wave proxy.
One such mechanism has been proposed by \cite{Rae2017}, whereby localized compressional 
ULF waves modulate the loss cone, bringing it within reach of a wider range of equatorial 
pitch angles for energetic particles. 
Thus, although the correlation between ULF waves and the VLF proxy that we found is mostly global,
it is possible that such local effects within more general ULF wave production events
may affect precipitation.
Indeed, the nature of storm-time ULF waves \citep[e.g.,][]{Anderson1990, Dai2015} 
which might be associated with precipitation is an important topic for future study. 
That is not to say that chorus waves do not exist 
during some times when precipitation is observed by POES; indeed satellite measurements have, 
on several occasions, shown chorus waves in correlation with those inferred using the 
precipitation VLF proxy \citep{Li2013POES, Li2016}. Rather, the precise level of VLF wave power 
and the spatial and temporal configuration of the inferred VLF waves may be modified by the 
effects of ULF wave-driven precipitation on the proxy. 
A further complicating factor is the degradation of the POES MEPED 
proton detectors \citep[e.g.,][]{galand2000, Odegaard2016}, 
whose directional dependence may affect
the L15 VLF proxy.

Overall, our results show a rather remarkable correspondence between the  VLF 
chorus wave proxy \citep{Li2013POES}, as applied in L15, and the distributions of global ULF wave power
in all 33 storms where such a comparison was feasible.
We note similar results by \cite{Katsavrias2015}, who examined only a single EA and IA 
event, and that of \cite{Ma2018} but without a discussion on morphological similarities.
This not only reveals further 
challenges with identifying the dominant and causative agents of radiation belt electron 
acceleration during storm time events, but also suggests the possibility of a much closer 
relationship between VLF waves and ULF waves in the inner magnetosphere than has previously 
been thought. Of course, it is possible that the reported correspondence between the VLF wave 
proxy and the observed ULF wave power is explained as a result of direct ULF wave driven 
precipitation polluting the VLF wave proxy. More work is needed to examine these 
inter-relationships further, but certainly our results suggest that care should be taken 
when using the POES precipitation measurements as a mechanism to derive a proxy for 
VLF wave amplitudes in the inner magnetosphere.

\acknowledgments
SD was supported by the Canadian Space Agency Geospace Observatory
(GO) Canada program. IRM is supported by a Discovery Grant from Canadian NSERC.
We thank Dr Wen Li for processed POES Chorus wave proxy data, as well as for her 
useful comments on the manuscript. The data in this letter are available at
https://osf.io/xdpf8/ ?view\_only$=$b0d7daa8b0ba488fae493d9ec1a729e4. 
All magnetometer data used are available from SuperMAG at 
http://supermag.jhuapl.edu/mag. 
For the magnetometer data we gratefully thank: USGS,
Jeffrey J. Love; CARISMA, PI Ian Mann; AARI, PI Oleg Troshichev; GIMA; 
The institutes who maintain the IMAGE magnetometer array, PI Eija Tanskane; 
AUTUMN, PI Martin Connors; DTU Space, PI Dr. Rico Behlke; CANMOS; 
THEMIS, PI Vassilis Angelopoulos.

\listofchanges

\end{document}